\documentclass{aa}  
\usepackage{natbib}
\usepackage{graphicx}
\usepackage[varg]{txfonts}
\usepackage{hyperref}
\usepackage{longtable}
\usepackage{caption}
\usepackage{subcaption}
\begin{document} 

        \title{Environmental effects on nearby debris discs}
        \author{A. M. Heras
                        \inst{1}
                        \and C. Eiroa\inst{2} 
            \and C. del Burgo\inst{3,4} 
            \and J. P. Marshall\inst{5}
            \and B. Montesinos\inst{6}}
    \institute{Directorate of Science, European Space Research and Technology Centre (ESA-ESTEC), Keplerlaan 1, 2201 AZ Noordwijk, The Netherlands 
        \\
                        \email{ana.heras@esa.int}
        \and
            Private Researcher, c/ Pablo Casals 20, 28011, Madrid, Spain
    \and
        Instituto de Astrofísica de Canarias, Vía Láctea S/N, La Laguna, E-38200, Tenerife, Spain
    \and
        Departamento de Astrofísica, Universidad de la Laguna, La Laguna, E-38200, Tenerife, Spain
    \and
        Academia Sinica Institute of Astronomy and Astrophysics, 11F of AS/NTU Astronomy-Mathematics Building, No.1, Section 4, Roosevelt Road, Taipei 106216, Taiwan
    \and
        Centro de Astrobiología (CAB) CSIC-INTA, ESAC Campus, Camino Viejo del Castillo s/n, E-28692, Villanueva de la Cañada, Madrid, Spain
                }
   \date{Received month day, year; accepted month day, year}
   \abstract
   {}
   {We investigate the influence of the interstellar medium (ISM) on debris discs using a statistical approach. We probe the effect of the ISM on debris disc occurrence rates and on the morphologies of the discs.}
   {We used results from the \textit{Herschel} Space Observatory DUNES and DEBRIS surveys of 295 nearby FGK dwarf stars imaged at 100\,$\mu$m and 160\,$\mu$m. Most of the 48 debris discs in this sample have small optical depths, making them more likely to be affected by the ISM compared to optically thick discs. Since the stars in our sample are located within the Local Interstellar Cloud, we can infer that their debris discs encounter similar conditions. This allows us to use the stellar space velocity, in particular the $U$ component, as a single indicator of the forces that can act on the debris disc dust grains when they interact with the ISM. Because older stars show a larger dispersion of space velocity values, we investigated the impact of the debris disc ages on our results.}
   {The observed debris disc occurrence rates seem to depend on the stellar space velocities, as expected under the hypothesis that stars with higher space velocities have a higher probability of losing their circumstellar dust. The percentage of sources with debris discs in our sample reaches a maximum of $\approx 25$\% for stars with low space velocity component values,  $|U_{\mathrm{rel}}|$, relative to the local ISM, and decreases monotonically for larger $|U_{\mathrm{rel}}|$ values down to the 10\% level. A decrease in the average disc fractional luminosity as a function of $|U_{\mathrm{rel}}|$ is also observed. These dependences do not disappear after accounting for the reported higher dispersion of $U$ values with age. In extended discs, the impact of the ISM could also explain the links observed between the stellar space velocities and the debris disc's projected ellipticities, position angles, and radii. The fractional luminosities of the debris discs appear to be correlated with their position angles, suggesting that the effect of the ISM on the dust content depends on the disc orientation. Although these indications may not be fully conclusive on their own, they collectively reinforce the hypothesis that the ISM influences the occurrence rates and morphologies of debris discs, thereby motivating additional research on the impact of the environment.  } 
   {}
   \maketitle
%
\nolinenumbers
\section{Introduction}
Debris discs are a natural outcome of planet formation processes. They consist of sub-micrometre- to millimetre-sized particles whose presence is explained by a continuous replenishment caused by collisions of planetesimals \citep[e.g.][]{Wyatt2008}. Dust grains absorb, scatter, and polarise light in the UV and visible wavelengths and re-emit most of the radiation they absorb through non-thermal and thermal mechanisms in the infrared and sub-millimetre. Consequently, the dust of debris discs can be observed by imaging their scattered light \citep{Schneider2014, Esposito2020, Ren2023} and/or their thermal emission \citep{Aumann1984,Matra2018a, Marshall2021}. Debris discs are generally depleted in gas, but high abundances of CO gas, probably also generated by second-generation processes when planetesimals collide, have been observed in some discs \citep[e.g.][]{Moor2017, Kral2020}.

Observations of mid- and far-infrared excesses with respect to stellar spectral energy distributions (SEDs) have provided a vast amount of information on the occurrence rates and structures of debris discs around main-sequence stars. The first discovery of a debris disc around a star was that of Vega by \citet{Aumann1984} using the Infrared Astronomical Satellite (IRAS). Their study led to a number of surveys with the Infrared Space Observatory (ISO) \citep{Habing2001} and the \textit{Spitzer} space observatory \citep{Werner2004}. \textit{Spitzer} showed that 16\% of solar-type FGK stars have debris discs \citep{Trilling2008}. Demographic studies with the \textit{Spitzer} Multiband Imaging Photometer (MIPS) at 24\,$\mu$m traced the decline with time of infrared excesses around main-sequence stars, with occurrence rates ranging from 50\% in the Beta Pictoris moving group (age of $\approx$ 20 Myr) to 6.5\% in Praesepe (age of around 800 Myr; \citealt{Chen2020}). \textit{Spitzer} observations have shown that excesses at 70\,$\mu$m decreased with age in a manner similar to the 24\,$\mu$m observations, which was consistent with the collisional evolution of their parent bodies \citep[see][for A and G stars, respectively]{Su2006, Kains2011}. However, at a given age, there is a large variation in the amounts of infrared excess present in the observed stars. The Photodetector Array Camera and Spectrometer (PACS) \citep{Poglitsch2010} and the Spectral and Photometric Imaging Receiver (SPIRE) \citep{Griffin2010} on board the \textit{Herschel} Space Observatory \citep{Pilbratt2010} enabled the observation of debris discs at far-infrared wavelengths with much higher sensitivity and spatial resolution. Using mainly 100 and 160\,$\mu$m PACS observations obtained as part of the Open Time Key Programmes (OTKPs) DUNES \citep{Eiroa2013} and DEBRIS \citep{Matthews2010}, \citet{Montesinos2016} determined the incidence of debris discs around FGK stars in the solar neighbourhood. Based on a sample of 105 such stars within a distance of 15 pc, they derived an overall debris disc incidence rate of 0.22$^{+0.08}_{-0.07}$, with 0.26$^{+0.21}_{-0.14}$ for F stars, 0.21$^{+0.17}_{-0.11}$ for G stars, and 0.20$^{+0.14}_{-0.09}$ for K stars (uncertainties correspond to the 95\% confidence level). The lowest observed values of the fractional luminosity were around $L_{\mathrm{d}}/L_{*} \le 4.0 \times 10^{-7}$. This is similar to the Solar System dust fractional luminosity estimated by \citet{Poppe2019} of $5 \times 10^{-7}$. On the other hand, \citet{Vitense2012} calculated the fractional luminosity of the Edgeworth-Kuiper Belt to be $1.2 \times 10^{-7}$, about three times lower than the minimum value in our sample \citep[an extended overview of the Solar System debris disc in the context of stellar debris discs can be found in][]{Horner2020}. Subsequently, \citet{Sibthorpe2018} analysed the debris disc occurrence in the \textit{Herschel} OTKP DEBRIS, an unbiased survey comprising the nearest $\approx90$ stars of each spectral type from A to M. They detected excess emission from a debris disc around 47 of the 275 F-K stars of the survey, for a detection rate of $17.1^{+2.6}_{-2.3}$ per cent, with lower rates for later spectral types. 

Departures from axisymmetry in debris discs have been interpreted as being due to gravitational sculpting by a planet, gravitational perturbations from stellar flybys or companions, interactions between gas and dust, planetesimal collisions, and interactions with the interstellar medium \citep[ISM; see e.g.][and references therein]{Hughes2018}. The ISM is the matter and radiation located between stars, composed of gas (a mixture of charged particles, atoms, and molecules), dust grains, and cosmic rays. The ISM behaves as a plasma with a relatively small mean free path between collisions and is subject to pressure forces. The motion of dust grains in debris discs can be strongly perturbed by interactions with the surrounding ISM particles. Interactions between disc dust and the ISM have been proposed to explain the asymmetric shapes of debris discs observed in scattered light. \citet{Debes2009} found that the asymmetries in the colour and morphology of the HD 32297 disc could be explained by the star's motion into a dense and cold ISM cloud at a relative velocity of 15 km s$^{-1}$. They argued that supersonic ballistic drag could explain the morphology of the debris discs of HD 32297, HD 15115, and HD 61005. \citet{Pastor2017} also found that the observed morphology of the debris ring in HD 61005 could be explained by considering dust under the action of the ISM, although this author required a different value for the ISM gas density than that obtained by \citet{Debes2009}. \citet{Maness2009}, on the other hand, modelled the asymmetries observed on the HD 61005 disc considering the process initially proposed by \citet{Scherer2000} as the dust removal mechanism in the Solar System. In this process, neutral gas moving through the system can introduce secular perturbations to the orbits of bound grains, significantly affecting the disc morphology and even unbinding the grains from the system. Considering this mechanism, \citet{Maness2009} suggested that HD 61005 was likely embedded in a more typical warm, low-density cloud of the local interstellar medium (LISM), which introduced secular perturbations to dust grain orbits. The effect of neutral gas on the orbits of bound grains was also the process favoured by \citet{MacGregor2015} to explain their tentative observation of millimetre emission aligned with the asymmetric western extension of the HD 15115 scattered light disc. However, movement through neutral gas may not affect all debris discs in the same way. \citet{Marzari2011} modelled the evolution of debris discs under the effects of solar radiation pressure, Poynting–Robertson (PR) drag, and ISM flux, and showed that for optical depths, $\tau$, of 10$^{-3}$ the effects of the ISM wind on grains with radii in the range 1--10\,$\mu$m were almost negligible. This optical depth is typical of dense collision-dominated discs such
as that around Beta Pictoris. However, when the optical depth was smaller, down to $\tau$ of 10$^{-6}$, significant asymmetries appeared on the modelled disc density profiles, both in the parent body's plane and out of plane. Discs with such low optical depths are almost collisionless, and their dynamics can be dominated by transport mechanisms, as occurs in the Kuiper Belt. In these low-optical-depth debris discs, the influence of the ISM combined with the PR drag could cause some particles to quickly migrate towards the star, where they are destroyed or sublimated, and thus create a sink of dust. In another study, \cite{Wijnen2017} proposed a mechanism based on theoretical solutions to the Stark problem to explain the spin--orbit misalignment observed in some planetary systems. The Stark problem is also known as the accelerated Kepler problem and refers to the particle motion in a Keplerian potential subject to an additional constant force. Through the development of an N-body model, \cite{Wijnen2017} show that a protoplanetary disc embedded in an ambient flow changes its orientation because its angular momentum vector tends to align parallel to the relative velocity vector.

The Sun and nearby stars are embedded in the Local Bubble, a region of low-density gas that extends in all directions to roughly 500 to 250 pc with hydrogen column densities of $\log N(\rm{H I}) = 9.3$ \citep{Lallement1986, Lallement2003}; it is surrounded by dense molecular clouds, such as Taurus and Ophiuchus. The Local Bubble contains about $10^5$ stars (including those with known exoplanets and debris discs) embedded in the LISM, which contains hot, warm, and cold gas. The Local Bubble has been shaped by the supernova explosions and winds of massive stars in the Scorpio-Centaurus association and has been ionised and heated by radiation from hot stars and the Galactic UV background. \citet{Zucker2022} discovered that the majority of star-forming regions close to the Solar System are located on the Local Bubble's surface, and the young stars within these regions predominantly expand outwards in a direction perpendicular to the bubble's boundary. Within the Local Bubble, the Local Interstellar Cloud (LIC) is a warm, partly ionised cloud encircling the Sun, whose internal pressure and ram pressure (both functions of its density and velocity relative to the Sun) impact the heliosphere, as well as the bubble-like regions of other stars \citep{Frisch2011}. \citet{Crutcher1982} noted that the interstellar gas in the LISM flows roughly in the same direction, away from the centre of the Scorpio-Centaurus association. \cite{Redfield2008} determined the physical properties of the LISM clouds close to the Sun using an empirical dynamical model based on 270 radial velocity measurements for 157 sightlines towards nearby stars. In particular, they characterised 15 warm clouds located within 15\,pc of the Sun, each with a different velocity vector. They found that the vector solutions for all clouds have similar directions, suggesting that there is a common history or a dynamical driver for all warm LISM clouds. \citet{Puspitarini2012} carried out a 3D mapping programme of the LISM and concluded that the expanding shells are difficult to reconcile with simple geometries. On the other hand, \citet{Gry2014} re-examined the LISM sightlines to stars within $d \leq$ 100\,pc and proposed a new structure for the LISM consisting of a single monolithic cloud that surrounds the Sun in all directions and accounts for most of the matter present in the first 50\,pc around the Sun. This cloud fills the space around us out to about 9\,pc in most directions, although its boundary is very irregular, possibly with a few extensions of up to 20\,pc. Still, the properties of dust in the LISM are poorly understood because the low column densities of dust towards nearby stars induce little photometric reddening, preventing the direct detection of the grains. To overcome this problem, stellar polarimetry has been used \citep{Marschall2016, Cotton2016, Cotton2017}. For example, \citet{Cotton2019} carried out multi-wavelength aperture polarimetry measurements of seven bright stars chosen to probe interstellar polarisation near the edge of the Local Bubble. Their results show that the large particles had either been shocked or swept up by past supernovae, so smaller grains define that region. Recently, \citet{Yeung2024}, using SRG/eROSITA observations, created a detailed map of the constituent of the soft X-ray background in the western Galactic hemisphere, focusing on the local hot bubble. 

\citet{Nordstrom2004} provided new determinations of age and kinematics, among other parameters, for a complete, magnitude-limited, and kinematically unbiased sample of 16\,682 nearby F and G dwarf stars. Their analysis showed a slow increase in the random space velocities with age, attributed to the heating of the galactic disc by massive objects such as spiral arms or giant molecular clouds. In particular, these authors showed that the dispersion of $U$ increases with age, so the older the star, the higher the probability that its $|U|$ is greater than 20 km\,s$^{-1}$. Generally, debris disc occurrence rates have been observed to decline with age. \citet{Sierchio2014} studied a sample of 238 F4–K2 stars with \textit{Spitzer} observations and found a significant reduction in the incidence of excesses at $70~\mu$m, from 22.5\% $\pm$ 3.6\% in stars younger than 5 Gyr to approximately 4\% in older stars. The proposed reason for this trend is that debris discs evolve collisionally over time \citep{Lohne2008, Lohne2012, Schuppler2014}, following passive disc evolution. However, \citet{Sierchio2014} also identified more excesses for the oldest stars than predicted by passive evolution models, suggesting  that the oldest debris discs could be activated by late-phase dynamical activity. \citet{Montesinos2016} also find indications that as stars get older, there is a slow erosion of the mass reservoir and dust content in the disc, resulting in a decrease in fractional luminosity. The reduction in dust content with age, established by theory and suggested by observations, and the connection between age and space velocity make it challenging to disentangle the contribution of each agent to the disc occurrence and fractional luminosities. In this work we aim to determine how to untwine the influence of these two factors. Our underlying assumption is that the enhanced ISM effect on debris discs that have higher space velocities is an important additional cause of dust depletion, which can operate in parallel or subsequent to collisional decay.

Distinguishing the effect of kinematics from that of other agents is not straightforward. As a recent example, \citet{Winter2020} proposed that the architecture of planetary systems strongly depends on local stellar clustering in position-velocity phase space. However, shortly afterwards, \citet{Mustill2021} debunked this claim, demonstrating that the preference for hot Jupiter hosts to be in high-density regions reflected an age bias, since younger stars are on average on colder orbits and hot Jupiters are more common around younger stars. 

While previous work has focused on the detailed analysis and modelling of individual objects, in this study we investigated the influence of the ISM on the occurrence rates and morphologies of debris discs in the solar neighbourhood using a statistical analysis. The study of the ISM effects on debris discs has often been hampered by uncertainties on the ISM properties at the locations of the stars under study. To avoid this limitation, we based our analysis on the DUNES debris disc sample \citep{Eiroa2013, Montesinos2016} combined with the DEBRIS sample selection made by \citet{Sibthorpe2018}, which represent a unique set of nearby objects located in the LIC region. They are optically thin debris discs, which are much more susceptible to being influenced by the LISM than optically thick discs \citep{Marzari2011}. Furthermore, observations conducted in the wavelength range of \textit{Herschel}/PACS trace dust grains that may be influenced by mechanisms different from those affecting the larger dust grains probed in Atacama Large Millimeter/submillimeter Array (ALMA) detections \citep[e.g.][]{MacGregor2019} and within timescales compatible with the crossing of the Local Bubble by these stars. 

In Sect. 2 we describe the data used in our study. In Sect. 3 we present our results on the dependence of debris disc occurrence rates on the star host space velocities. This is followed by the results of our analysis of the large-scale morphology of our extended debris discs and possible correlations with the magnitude and orientation of the host star space velocities in the LISM. In Sect. 4 we discuss our findings, and in Sect. 5 we provide our conclusions. 
\begin{figure}
        \includegraphics[width=\hsize]{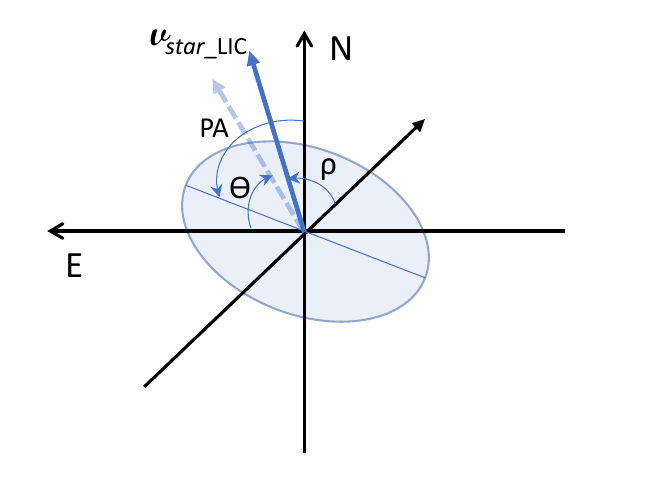}
                \caption{Geometry of the debris disc, showing the north, east, and radial directions on the sky and the disc PA. The vector of the space velocity of the host star with respect to the LIC, $v_{\mathrm{star\_LIC}}$, is represented by a blue arrow, which has angle $\rho$ with respect to the radial direction. Angle $\theta$ is the angle of the sky projection of $v_{\mathrm{star\_LIC}}$ with respect to the east sky direction.}
                \label{geometry}
\end{figure}
\section{Data}
In this work we used results from the DUNES and DEBRIS surveys proposed as OTKPs to the \textit{Herschel} Space Observatory, which obtained images at 100\,$\mu$m and 160\,$\mu$m of nearby stars, complemented in some cases with observations at 70\,$\mu$m \citep{Matthews2010, Eiroa2013, Sibthorpe2018, Marshall2021}. The DUNES sample is sensitivity- and background-limited, and contains FGK solar-like stars located at $d \leq 20$ pc. The \citet{Sibthorpe2018} sample is volume-limited and consists of FGK solar-like stars. We excluded the six cold discs in the DUNES sample identified in \citet{Eiroa2011}, \citet{Marshall2013}, and \citet{Krivov2013} because they have only been observed at 160\,$\mu$m and are still controversial. The maximum distance of the combined samples is 25 pc. These stars are embedded in the LISM at distances at which the LISM parameters have been determined. Consequently, we can consider that their debris discs are all affected by alike ISM conditions, which supports our statistical approach.

For most stars, we derived the galactic space velocity components $U$, $V$, and $W$ in the heliocentric rest frame following \cite{Johnson1987}, starting from the target coordinates, the radial velocities, and the proper motions given in the \textit{Gaia} DR3 catalogue \citep{Gaia2016,Gaia2021,Seabroke2021}. Positive values for $U$, $V$, and $W$ respectively indicate velocities towards the galactic centre ($l = 0\degr$, $b = 0\degr$), the galactic rotation, and the North Galactic Pole ($b = 90\degr$). For the brightest targets that do not have radial velocities in the \textit{Gaia} DR3 catalogue, we adopted space velocity values given in the literature. The full sample with the information on the presence of debris discs, their dust fractional luminosity, detected planets, ages, and their heliocentric space velocities is given in Table A.1. 
According to \citet{Redfield2008}, the heliocentric velocity vector of the LIC has a magnitude of 23.84 km\,s$^{-1}$, with a direction in galactic coordinates given by $b_{0} = -13.5^{\degr}$  and $l_{0}= 187^{\degr}$. This corresponds to a heliocentric space velocity $U =-22.8$~km\,s$^{-1}$, $V=-4.0$~km\,s$^{-1}$, and $W =-5.6$~km\,s$^{-1}$. Based on these values, we calculated the space velocity $U_{\mathrm{rel}}$, $V_{\mathrm{rel}}$, and $W_{\mathrm{rel}}$ of each star relative to the LISM. The values for the heliocentric velocity vector of the LIC were confirmed by \citet{McComas2012} using in situ measurements by IBEX  of the motion of individual interstellar particles. 

Some of the detected debris discs are resolved in the \textit{Herschel} PACS images \citep{Eiroa2013,Montesinos2016}. Because the relatively low resolution of the images prevents a detailed study of the debris disc morphologies, we searched for correlations between some disc parameters (see Table \ref{table:ext}) and the stellar space velocities relative to the LISM. The debris discs' position angles (PAs) on the sky (measured from north to east) and the projected elliptical diameters (major, $a$, and minor, $b$) of the 3$\sigma$ contours have been taken from \citet{Eiroa2013} or measured for this work. The uncertainty for the PA values is $\pm5\degr$. These morphology values are based on the 100\,$\mu$m images, except for two targets, for which only 70$\mu$m data were available, as indicated in Table~\ref{table:ext}. The projected ellipticities of the discs, $\epsilon$, have been derived as $\epsilon = 1-b/a$. We calculated $\rho$, the angle between the radial direction and the relative velocity of the host star with respect to the LIC, $v_{\mathrm{star\_LIC}}$. Similarly, we calculated $\theta$, the angle of the sky projection of $v_{\mathrm{star\_LIC}}$ with respect to the east direction of the sky. The results are shown in Table \ref{table:ext}, to which we have added the estimated debris disc radii derived by \citet{Marshall2021}, when available, or calculated following the same method. 

Figure~\ref{geometry} represents the geometry with the angles used in this study. We did not consider the stellar rotation axis in our work. Although \citet{Greaves2014} showed that the debris discs were well aligned with the stellar rotation axis in a sample of 11 objects, those debris discs had higher optical depths, making them less vulnerable to the environment. Therefore, a larger study of debris discs with smaller optical depths is required to establish more general conclusions about stellar axis and debris disc alignments. 
\begin{table*}
\caption{Extended debris discs.}             
\label{table:ext}       
\centering                          
\begin{tabular}{c c c c c c c c c}        
\hline\hline                 
HIP & PA ($\degr$) & Size $a$ ($"$) & Size $b$ ($"$) & $\epsilon$ & $\rho$ ($\degr$) & $\theta$ ($\degr$) & disc $R$ (au) \tablefootmark{1} & $L_{\mathrm{dust}}/L_{*}$ \\   
\hline                        
544 & 102& 18 & 16 & 0.111 $\pm$ 0.001 & -45.63 & -81.16 & $27.3_{-2.8}^{3.4}$ &$4.80 \times 10^{-5}$ \\ 
\\[-0.5em] 
1368 & 10& 12 & 7 & 0.417 $\pm$ 0.011 & -73.76 & 18.26 & $54.5_{-12.4}^{6.4}$ &$9.80 \times 10^{-5}$ \\ 
\\[-0.5em] 
7978 & 51& 39 & 28 & 0.282 $\pm$ 0.002 & 44.60 & -102.21 & $81.1_{-1.3}^{0.8}$ &$3.10 \times 10^{-4}$ \\ 
\\[-0.5em] 
8102\tablefootmark{2} & 110& 11 & 10 & 0.091 $\pm$ 0.002 & -59.68 & 173.27 & $6.4_{-1.1}^{0.7}$ &$6.10 \times 10^{-6}$ \\ 
\\[-0.5em] 
13402 & 50& 19 & 14 & 0.263 $\pm$ 0.004 & 87.92 & -72.61 & $25_{-2}^{2}$ &$1.70 \times 10^{-5}$ \\ 
\\[-0.5em] 
14954 & 30& 18 & 14 & 0.222 $\pm$ 0.003 & -89.46 & -56.72 & $76.2_{-5.9}^{5.7}$ &$4.20 \times 10^{-6}$ \\ 
\\[-0.5em] 
16537\tablefootmark{2} & 0& 25 & 21 & 0.160 $\pm$ 0.002 & -82.88 & -161.20 & $16.5_{-0.3}^{0.4}$ &$5.40 \times 10^{-5}$ \\ 
\\[-0.5em] 
16852 & 0& 8.5 & 7.7 & 0.094 $\pm$ 0.003 & 81.21 & -124.09 & $46_{-4}^{4}$ &$8.37 \times 10^{-6}$ \\ 
\\[-0.5em] 
17420 & 50& 3.5 & 2.5 & 0.286 $\pm$ 0.056 & 59.36 & -8.85 & $60_{-5}^{5}$ &$9.20 \times 10^{-6}$ \\ 
\\[-0.5em] 
17439 & 105& 26 & 14 & 0.462 $\pm$ 0.006 & 27.57 & 33.95 & $49.3_{-2.1}^{4}$ &$8.10 \times 10^{-5}$ \\ 
\\[-0.5em] 
19893 & 40& 10 & 6 & 0.400 $\pm$ 0.012 & 17.11 & -40.05 & $80.6_{-2.6}^{4.8}$ &$1.90 \times 10^{-5}$ \\ 
\\[-0.5em] 
22263 & 5& 21 & 16 & 0.238 $\pm$ 0.003 & 77.61 & -14.00 & $29_{-2.4}^{4.7}$ &$2.90 \times 10^{-5}$ \\ 
\\[-0.5em] 
32480 & 107& 34 & 20 & 0.412 $\pm$ 0.004 & -30.37 & 69.93 & $93.9_{-2.4}^{3}$ &$6.90 \times 10^{-5}$ \\ 
\\[-0.5em] 
51502 & 0& 16 & 16 & 0.000 $\pm$ 0.000 & -89.18 & 24.18 & $54.5_{-7.7}^{6.9}$ &$1.30 \times 10^{-5}$ \\ 
\\[-0.5em] 
61174 & 115& 16.6 & 12.9 & 0.223 $\pm$ 0.004 & 65.89 & -150.96 & $85_{-3}^{20}$ &$2.17 \times 10^{-5}$ \\ 
\\[-0.5em] 
62207 & 130& 24 & 14 & 0.417 $\pm$ 0.006 & 5.89 & 167.93 & $63.9_{-4.9}^{5.3}$ &$2.10 \times 10^{-5}$ \\ 
\\[-0.5em] 
64924 & 65& 17.4 & 8.7 & 0.500 $\pm$ 0.010 & 84.31 & -117.61 & $47_{-1}^{1}$ &$2.44 \times 10^{-5}$ \\ 
\\[-0.5em] 
72848 & 75& 21 & 12 & 0.429 $\pm$ 0.007 & -52.94 & 176.85 & $33_{-9}^{9}$ &$2.80 \times 10^{-6}$ \\ 
\\[-0.5em] 
85235 & 0& 18 & 16 & 0.111 $\pm$ 0.001 & -49.29 & -144.97 & $33.3_{-2.8}^{6.6}$ &$2.00 \times 10^{-5}$ \\ 
\\[-0.5em] 
88745 & 72& 17.7 & 12.8 & 0.277 $\pm$ 0.004 & 46.60 & -135.64 & $130.3_{-4.2}^{5.7}$ &$1.11 \times 10^{-5}$ \\ 
\\[-0.5em] 
107649 & 125& 17.5 & 8 & 0.543 $\pm$ 0.012 & 83.59 & -115.98 & $107.6_{-1.8}^{2.8}$ &$1.00 \times 10^{-4}$ \\ 
\\[-0.5em] 
114948 & 0& 5 & 5 & 0.000 $\pm$ 0.000 & 55.60 & -139.88 & $44_{-4}^{4}$ &$2.70 \times 10^{-5}$ \\ 
\\[-0.5em] 
\hline         
\end{tabular}
\tablefoot{
\tablefoottext{1}{From or derived as in \citet{Marshall2021}; }
\tablefoottext{2}{only observed at 70\,$\mu$m}}
\end{table*}
\section{Results}
\subsection{Debris disc occurrence rate}\label{section:occurrence}
\begin{figure}
\resizebox{\hsize}{!}
        {\includegraphics[trim= 0 60 0 0, clip]{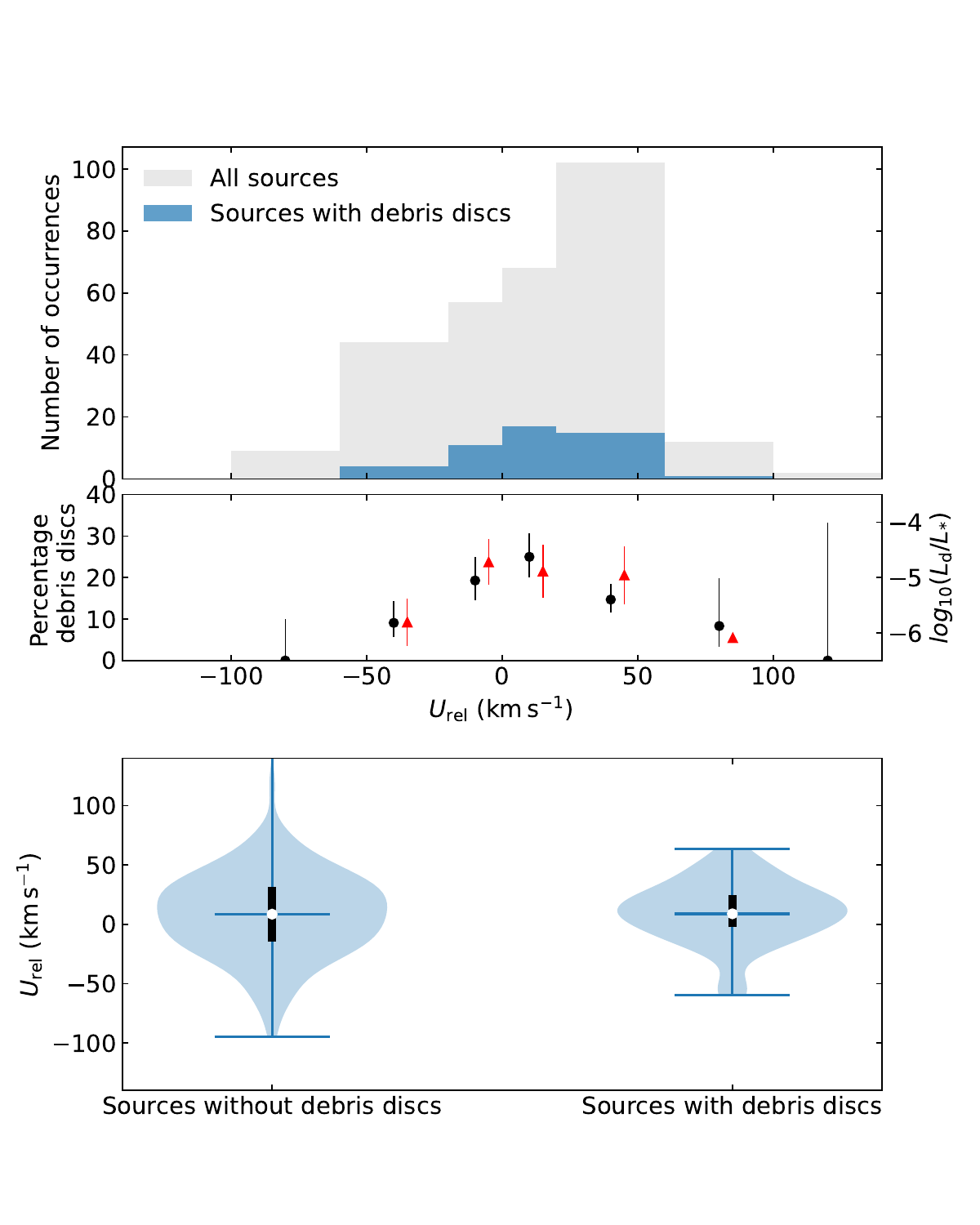}}
                \caption{Occurrence rate of debris discs as a function of space velocity component $U_{\mathrm{rel}}$. Top panel: Histograms showing the total number of stars (grey bars) and stars with debris discs (blue bars) as a function of $U_{\mathrm{rel}}$. Middle panel: Percentage of stars that have debris discs  (black dots) and the median of the logarithm of the disc fractional luminosity in each bin (red triangles) as a function of $U_{\mathrm{rel}}$. Bottom panel: Violin diagrams representing the probability density distribution of targets with and without debris discs as a function of $U_{\mathrm{rel}}$. The white dots show the median values, and the thick black lines indicate the interval between the first and third quartiles. The horizontal width of each blue-shaded violin area corresponds to the relative frequency of the corresponding objects as a function of $U_{\mathrm{rel}}$. To enhance visibility, the upper limit at $U_{\mathrm{rel}}=303$\,km\,s$^{-1}$ of the first violin is not included in the figure. We see that the probability density distribution as a function of $U_{\mathrm{rel}}$ is narrower for stars with debris discs than for stars without debris discs.}\label{histo}
\end{figure}

\begin{figure}
 \resizebox{\hsize}{!}
        {\includegraphics[trim= 0 40 0 0, clip]{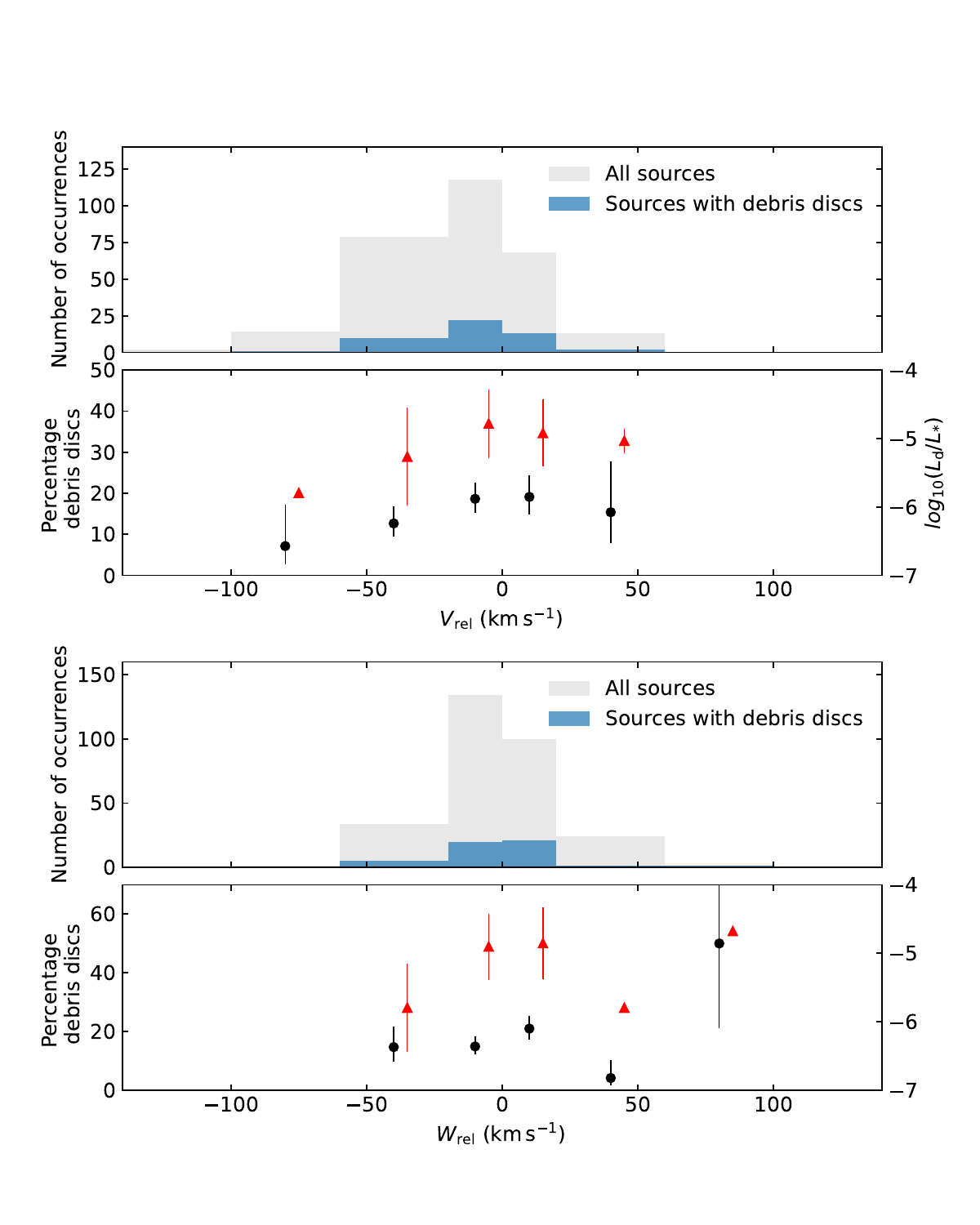}}
        \caption{Occurrences of stars that have debris discs as a function of space velocity components $V_{\mathrm{rel}}$ (top panel) and $W_{\mathrm{rel}}$ (bottom panel), as in the two upper panels of  Fig.~\ref{histo}.} 
        \label{histo_all}
\end{figure}

The first question we address is whether the occurrence of debris discs around the FGK stars of our sample is affected by the interaction with the LISM. Figure~\ref{histo} shows the distribution of the FGK stars considered in this study as a function of their space velocity $U_{\mathrm{rel}}$. The distribution of targets with detected debris discs is overplotted. The middle panel displays the percentage of stars that have debris discs (black dots) and the median of the logarithm of dust fractional luminosity of the discs in each bin (red triangles) as a function of $U_{\mathrm{rel}}$. For the disc occurrence percentages, the error bars correspond to the binomial proportion 1-$\sigma$ confidence interval. The percentage of sources with debris discs reaches a maximum of $\approx 25$\% for stars with low values of $|U_{\mathrm{rel}}|$, and decreases monotonically for larger $|U_{\mathrm{rel}}|$ values down to the 10\% level. A similar behaviour is seen for the disc fractional luminosities averaged per bin, suggesting that not only the frequency of debris discs, but also their dust content, decreases with $|U_{\mathrm{rel}}|$. In the bottom panel of Fig.~\ref{histo}, a violin diagram shows that the probability density distribution of targets with debris discs as a function of $U$ is narrower than the distribution of targets without debris discs. The median values are very similar, but the interval between quartiles, represented by the black lines, is smaller in the sources with debris discs. 
We assessed the statistical significance of this comparison by applying a Levene test, which evaluates the null hypothesis that the input samples are from populations with equal variances. In our case, the Levene test provides a p-value of 0.02, indicating a low probability that the variances with respect to $U_{\mathrm{rel}}$ of the population of targets with and without debris discs are the same. The observed decrease in the probability density distribution of debris discs with increasing space velocity, $|U_{\mathrm{rel}}|$, is consistent with a scenario in which the LISM interacts with debris discs to the extent that stars with larger space velocities have a higher probability of losing their circumstellar dust. This is supported by the decrease in average dust fractional luminosity with larger $|U_{\mathrm{rel}}|$. One limitation is that our sample includes a small number of stars with $|U_{\mathrm{rel}}|$ > 50\,km\,s$^{-1}$, reducing the significance of the results in that range. We also find that the interaction of the LISM with the debris discs is mainly defined by the $U$ component of the stellar velocity, as can be inferred from Fig.~\ref{histo_all}. The dependence of debris disc occurrence on the velocity component $V$ is less apparent and is not present in association with the velocity component $W$ when $W_{\mathrm{rel}} < 30\,\rm{km\,s}^{-1}$. However, for the small number of stars with $W_{\mathrm{rel}}$ greater than this value, the debris disc occurrence rate shows a strong decrease. A reason could be that these stars have travelled from a region with different conditions, possibly outside of the LISM, which has resulted in their debris discs being more ablated. 

\begin{figure}
    \resizebox{\hsize}{!}
           {\includegraphics{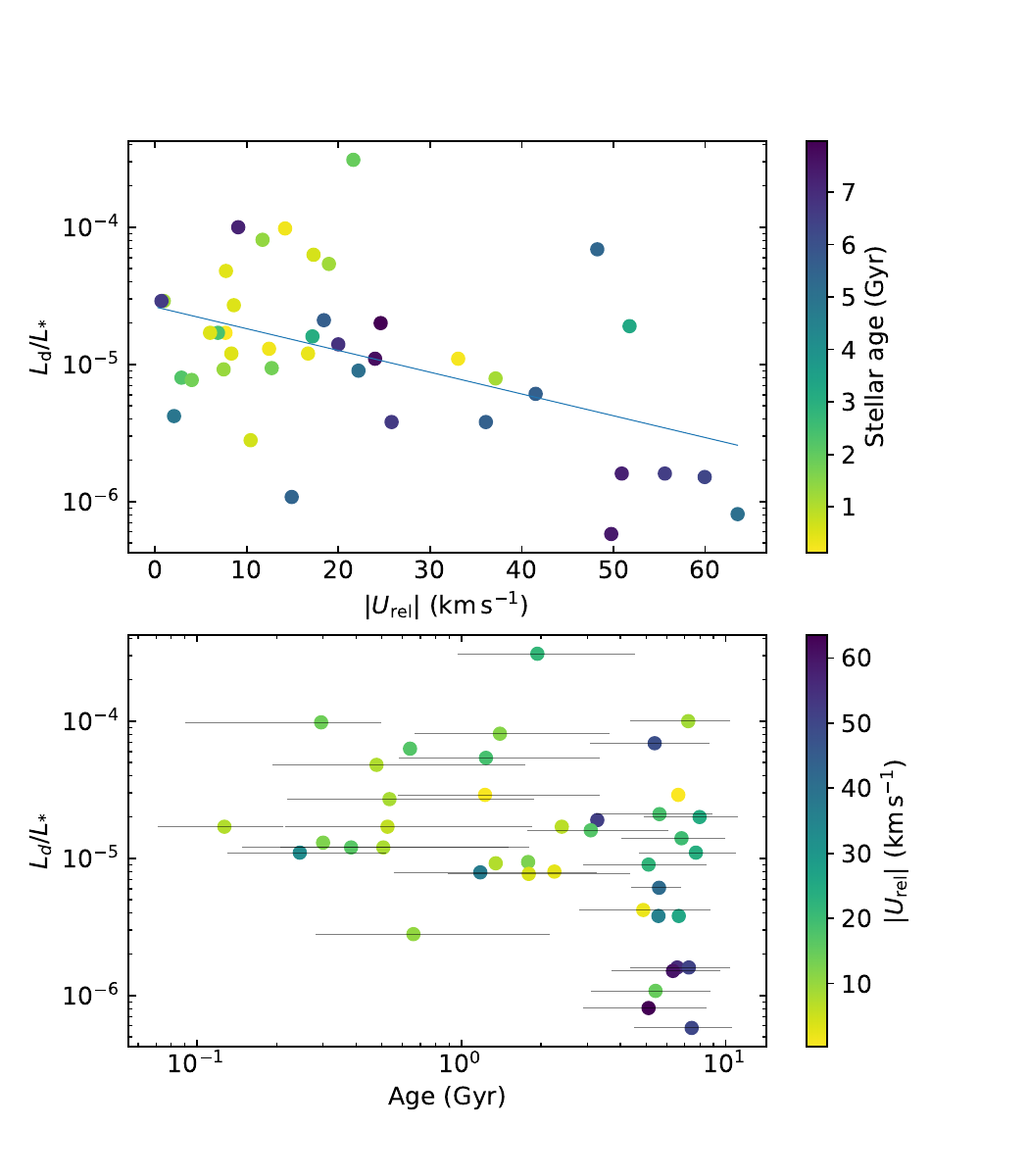}}
           \caption{Debris disc fractional luminosity vs. space velocity and age. Upper panel: Debris disc fractional luminosity as a function of space velocity, $|U_{\mathrm{rel}}|$, with colours from yellow to dark blue indicating young to old stars, respectively. The line is the result of linear regression fitting. Lower panel: Debris disc fractional luminosity as a function of stellar age, with colours from yellow to dark blue indicating increasing space velocity, $|U_{\mathrm{rel}}|$. }
           \label{uvel_Ld}
\end{figure}

 A complementary view of the dependence of the disc fractional luminosities, $L_{\mathrm{d}}/L_{*}$, with $|U_{\mathrm{rel}}|$ is shown in Fig. \ref{uvel_Ld}. The $L_{\mathrm{d}}/L_{*}$ uncertainties are about 10\% or less \citep{Yelverton2019}. As expected, the dust fractional luminosity shows a large scatter, since it depends on various factors intrinsic to the formation and evolution of debris discs and their planetesimals. However, and although the numbers are low, Fig. \ref{uvel_Ld} suggests that the amount of dust in our debris discs decreases with increasing $|U_{\mathrm{rel}}|$, as was already apparent in Fig. \ref{histo}. We applied Fisher's exact test to four groups consisting of targets with debris discs above or below $L_{\mathrm{d}}/L_{*} = 10^{-5}$ and above or below $|U_{\mathrm{rel}}| = 25$ km\,s$^{-1}$, and obtained a p-value of 0.006. That is, the observed distributions of $L_{\mathrm{d}}/L_{*}$ for objects with high or low $|U_{\mathrm{rel}}|$ are probably different. As indicated by the colours, which represent the stellar ages listed in Table A.1 and derived by \citet{Stanford-Moore} and by \citet{Montesinos2016}, the disc fractional luminosities also decrease with stellar age. However, this is valid mainly for stars younger than $\approx$3\,Gyr. In older stars, fractional luminosity values do not show a dependence on age, the trend with $|U_{\mathrm{rel}}|$ being the dominant factor. Fisher's exact test, taking four groups of targets with discs above or below $L_{\mathrm{d}}/L_{*} = 10^{-5}$ and ages greater than or lower than 3 Gyr yields a p-value of 0.11, greater than when considering $|U_{\mathrm{rel}}|$. A larger number of debris discs hosted by stars with high spatial velocities is certainly needed to confirm or reject this tentative trend. We do not observe similar trends for the components $V$ or $W$, which is consistent with the observed non-dependence of the debris disc occurrence rates on these components of the space velocity.
\begin{figure}
 \resizebox{\hsize}{!}
        {\includegraphics[trim= 0 20 0 0, clip]{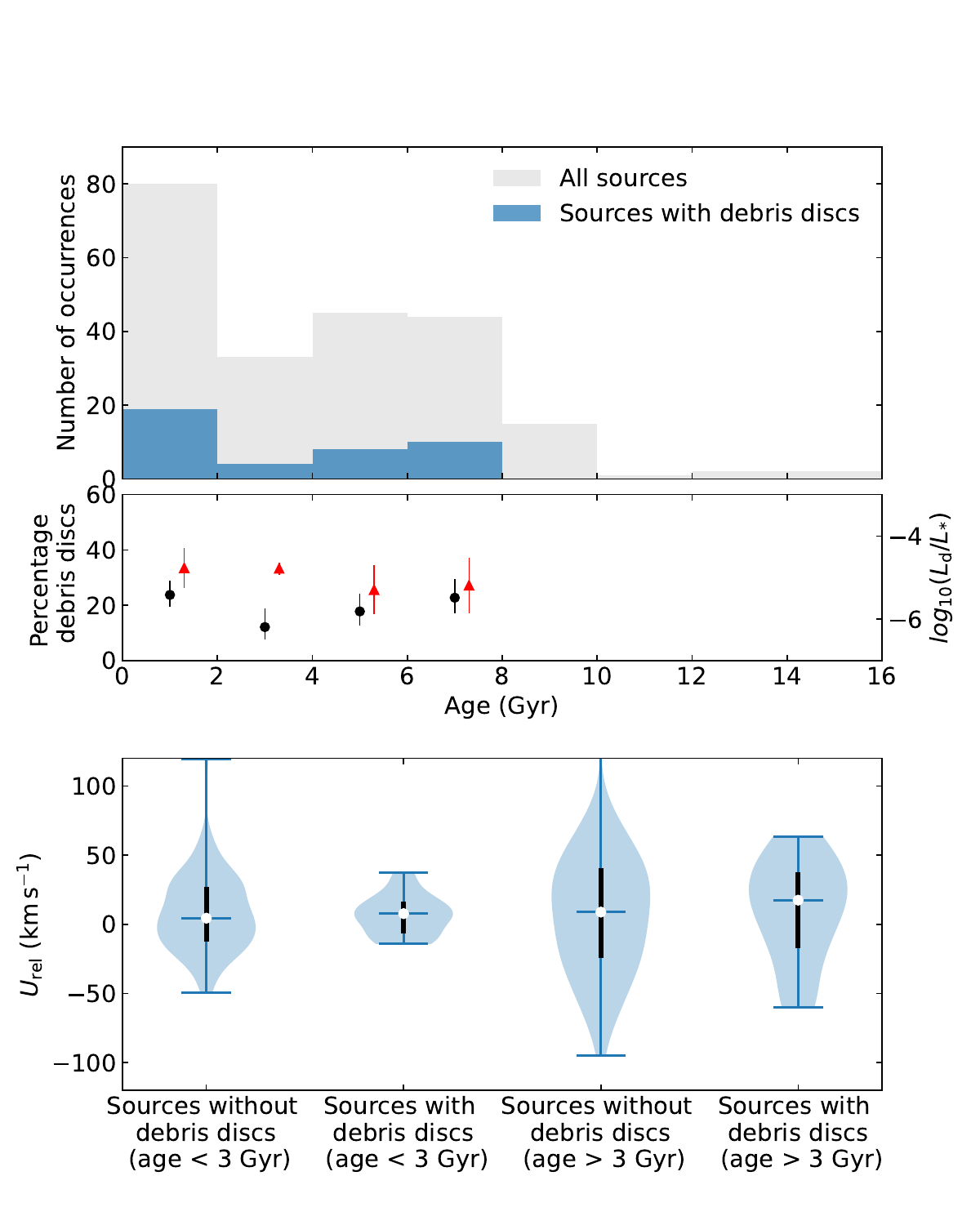}} 
        \caption{Occurrence rate of debris discs as a function of age. Top panel: Histograms showing the number of stars and stars with debris discs as a function of age. Middle panel: Percentage of stars that have debris discs (black dots) and the median of the logarithm of disc fractional luminosity in each bin (red triangles) as a function of age. Bottom panel: Probability density distributions for younger stars (age < 3 Gyr) and older stars (age > 3 Gyr), with and without debris discs, as a function of the space velocity, $U_{\mathrm{rel}}$. For clarity, the upper limit at $U_{\mathrm{rel}} = 303$\,km\,s$^{-1}$ of the third violin has not been included in the figure.}
        \label{U_age}
\end{figure}
\par
\begin{figure}
    \resizebox{\hsize}{!}
           {\includegraphics{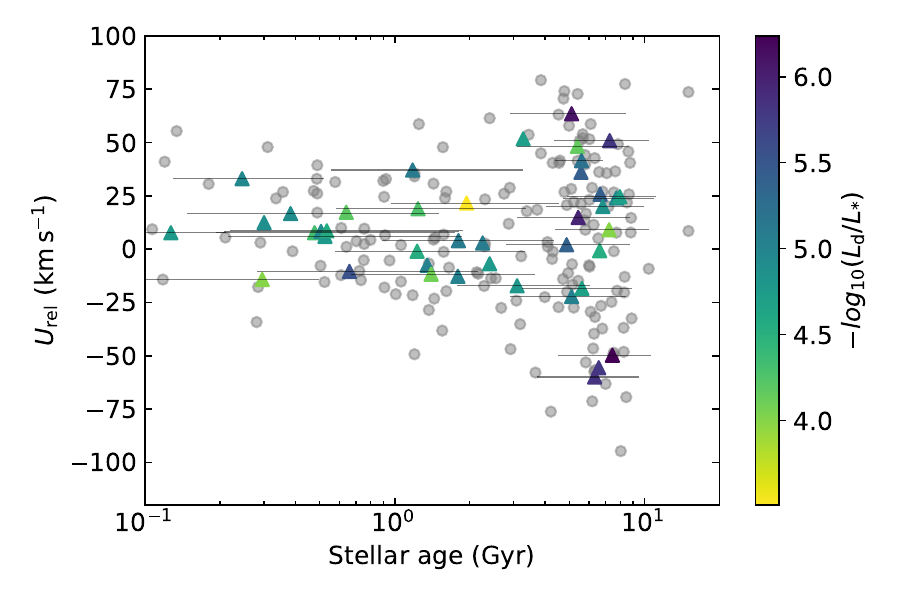}}
           \caption{Stellar space velocity, $|U_{\mathrm{rel}}|$, as a function of stellar age for the targets of our sample, without debris discs (grey dots) and with debris discs (triangles coloured from yellow to dark blue, from higher to lower fractional luminosity).} 
           \label{age_uvel}
\end{figure}

 We analysed the evolution of debris disc occurrence rates with age (top panel of Fig.~\ref{U_age}), with the caveat that there are still significant inaccuracies in the derivation of stellar ages, and these uncertainties may have an important effect in the search for trends. Our derived occurrence rates show only a slight decrease as a function of age for stars younger than 4\,Gyr, but this trend is discontinued for older stars, which is consistent with the findings of \citet{Sierchio2014}. The increase in space velocity $|U|$ with age found by \citet{Nordstrom2004} becomes apparent in the bottom panel of Fig.~\ref{U_age}, which displays the probability density distributions for younger stars (age < 3\,Gyr) and older stars (age > 3\,Gyr), with and without debris discs, as a function of space velocity $U_{\mathrm{rel}}$. In agreement with the results of \citet{Nordstrom2004}, the probability density distribution for older stars spreads to a wider range of $U$ values than the one for younger stars, both for sources with and without debris discs. However, for each age group, sources with debris discs still have smaller values of $|U_{\mathrm{rel}}|$ than sources without debris discs.  The Levene test gives a p-value of 0.014 when comparing the distributions with and without debris discs for stars younger than 3 Gyr, indicating a low probability that the input samples are from populations with equal variances.  Although there is likely an impact of the age of the stars on the debris disc occurrence rates, these findings support that the space velocity dependence and therefore the effect of the LISM is at least as important. On the other hand, the Levene test p-value that takes both the distributions with and without debris discs into account for stars older than 3 Gyr is 0.38, which is consistent with the two distributions having the same variance. To provide a more detailed view, we have plotted in Fig.~\ref{age_uvel} the distribution of $|U_{\mathrm{rel}}|$ versus stellar age for the targets in our sample. The increase in the space velocity dispersion with age becomes apparent after 1 Gyr, with the highest values of $|U_{\mathrm{rel}}|$ reached by stars older than 4 Gyr. Objects with debris discs follow this trend as well. However, debris discs appear preferably in objects with lower $|U_{\mathrm{rel}}|$, in agreement with the bottom panel of Fig.~\ref{U_age}. In stars with age\,>\,3\,Gyr, we additionally observe that five out of six discs with $L_{\mathrm d}/L_{*} < 3 \times 10^{-5}$ have $|U_{\mathrm{rel}}| > 40$\,km\,s$^{-1}$. That is, although the analysis of Fig.~\ref{U_age} (bottom planet) does not show that the debris disc frequency in stars older than 3 Gyr decreases with $|U_{\mathrm{rel}}|$, we see that their dust content does, becoming $|U_{\mathrm{rel}}|$ a differentiating factor for stars of similar age. 
 
\begin{figure}
  \resizebox{\hsize}{!}
        {\includegraphics[trim= 0 60 0 0, clip]{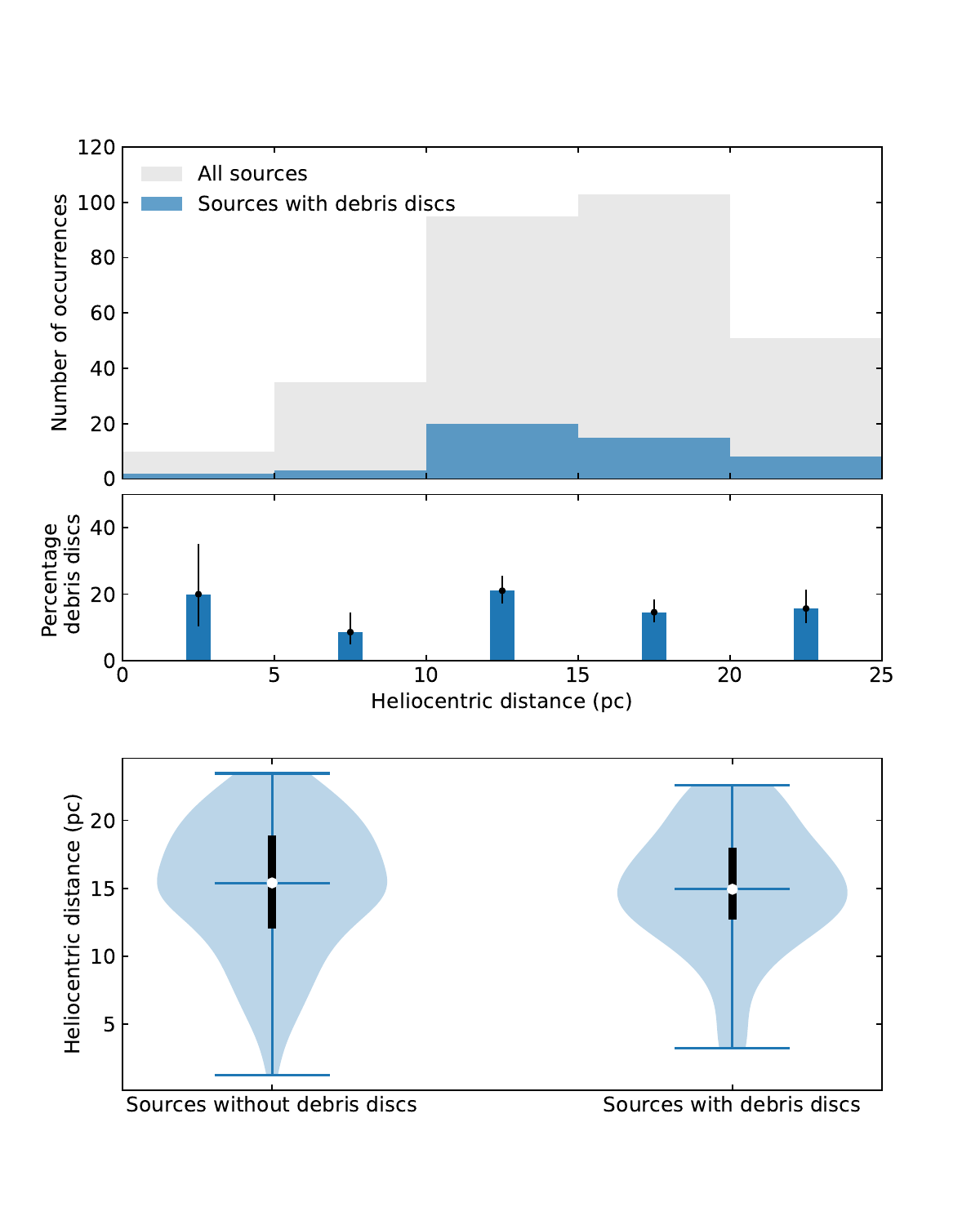}}
        \caption{Occurrence rate of debris discs as a function of heliocentric distance. Top panel: Histograms showing the total number of stars (grey bars) and stars with debris discs (blue bars) as a function of heliocentric distance. Middle panel: Histogram showing the percentage of stars that have debris discs as a function of heliocentric distance. Bottom panel: Probability distributions as a function of heliocentric distance (a more detailed explanation of the violin diagrams can be found in the caption of Fig.~\ref{histo}).}
    \label{histo_all_dist}
\end{figure}

\begin{figure}
    \resizebox{\hsize}{!}
        {\includegraphics{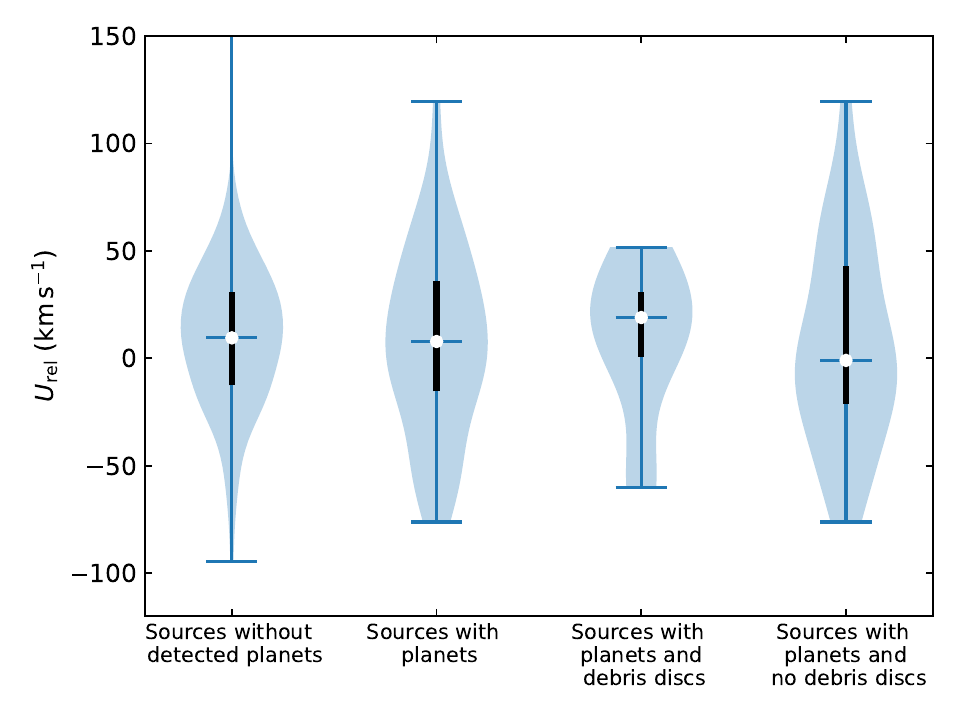}}
        \caption{Probability density distributions for sources without planets, with planets, with planets and debris discs, and with planets but no debris discs, as a function of the space velocity $U$. For clarity, the upper limit at $U=303$\,km\,s$^{-1}$ of the first violin has not been included in the figure.}
        \label{planets}
\end{figure}

Studies of the LISM have indicated that it is not completely uniform and that a certain structure can be inferred from the observations. Ultraviolet and optical spectra of interstellar gas along the lines of sight to nearby stars have been used to characterise the LISM as a set of discrete warm, partially ionised clouds, each with a different velocity vector, temperature, and metal depletion \citep[][and references therein]{Redfield2008}. In contrast, \citet{Gry2014} proposed a fundamentally different model to describe the LISM, consisting of a single continuous cloud with non-rigid flows filling the space out to 9\,pc from the Sun, with a lower limit of 1.3\,pc for the minimum cloud extent in any direction. With these different LISM views in mind, we derived the debris disc occurrence rate as a function of the distance from the Sun. Although the sample size of the nearest stars is insufficient for a statistically significant conclusion, Fig. \ref{histo_all_dist} suggests a slight tendency for a lower occurrence rate of debris discs in stars with distances between 5 and 10\,pc, as compared to those further away. This is supported by the violin distributions as a function of the heliocentric distance displayed in the bottom panel of Fig.~\ref{histo_all_dist}, which show that the probability density distribution of stars with debris discs is narrower at shorter distances than for stars without debris discs. This tentative difference could be explained by assuming that debris discs around stars at distances within 10\,pc are under stronger influence of the LISM because it is denser in this region, a scenario that is consistent with the model proposed by \citet{Gry2014}. An additional feature is that the occurrence rate decreases again for stars at distances greater than 14\,pc. This decrease could also be explained by the observations from \citet{Gry2014} that indicate there are a few directions in which the local cloud, which is denser than the Local Bubble, could extend up to about 20\,pc. However, it may also be associated with an increase in the lower limits of the observable dust fractional luminosity for the farthest stars, which is especially apparent for spectral types G and K \citep[see Fig. 8 in][]{Eiroa2013}. 
The search for correlations between the presence of planets and debris discs has been the subject of a number of studies. \citet{Wyatt2012} and \citet{Marshall2014b} found correlations between the presence of debris discs and planets, although their results were affected by the small number statistics for the sample with planets. More recently, \citet{Meshkat2017} also found such a correlation based on a high-contrast imaging survey of planets. These authors found that the frequency of giant planets with masses of 5–20 M$_{\rm Jup}$ and separations of 10–1000 au around stars with debris discs was 6.27\% compared to 0.73\% for the control sample of stars without discs, with a difference between these distributions at the 88\% level. A different conclusion was reached by \citet{Yelverton2020} using a larger sample than in previous works, which consisted of 201 stars known to host planets (mostly detected through RV) and a control sample of 294 stars without known planets. These authors found no evidence that the presence of RV planets significantly affected the fractional luminosities of debris discs, in agreement with the analysis performed by \citet{Moro2015}. In the context of these studies, we investigated to what extent our data show correlations between debris discs and planets and if these trends are affected by the influence of the ISM in which the host star moves. We can only draw tentative conclusions due to the small number of objects in our sample. Of the 295 FGK stars, only 41 have planets, and out of these 11 have a debris disc. In Fig.~\ref{planets}, the probability density distributions of these populations are represented as a function of the space velocity $U_{\mathrm{rel}}$. As can be seen, the sources with and without planets have similar distributions with respect to $U_{\mathrm{rel}}$. On the other hand, the distribution of sources with planets and debris discs appears to be narrower with respect to $U_{\mathrm{rel}}$ than the distribution of sources with planets but without debris discs. This tentative difference could have implications when looking for correlations between the properties of the debris discs and the presence of planets. Stars with planets and large space velocities might have lost part or most of their circumstellar dust as a consequence of the interaction with the ISM. That is, if present, the effect of the ISM would mask the dependence between the presence of planets and the occurrence and characteristics of debris discs. 
\subsection{Extended discs}
\begin{figure*}
\resizebox{\hsize}{!}{
         \includegraphics{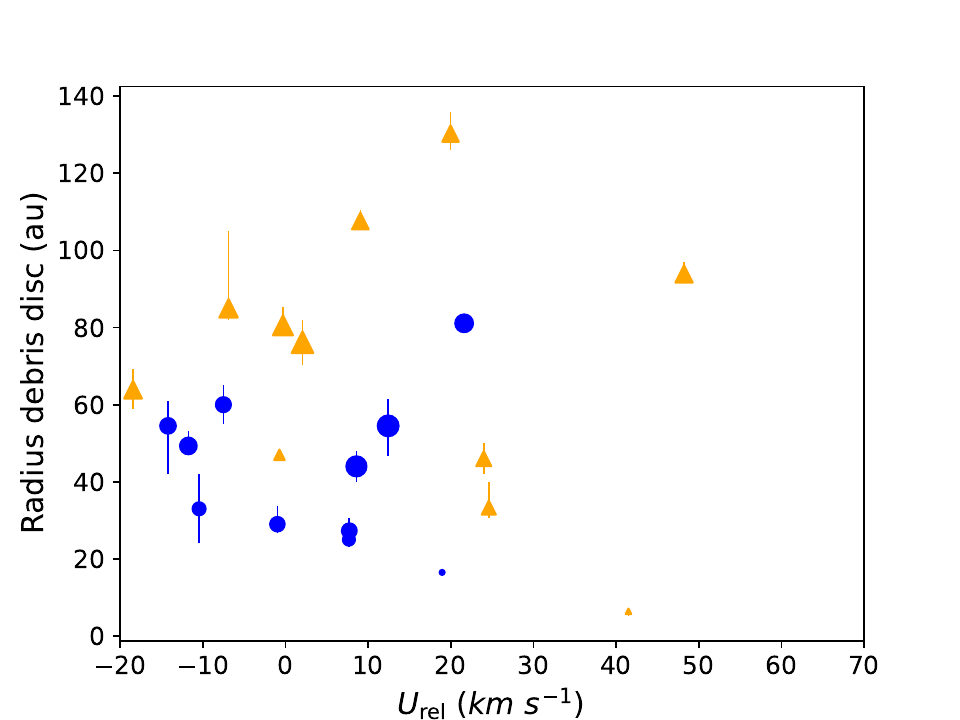}
         \includegraphics{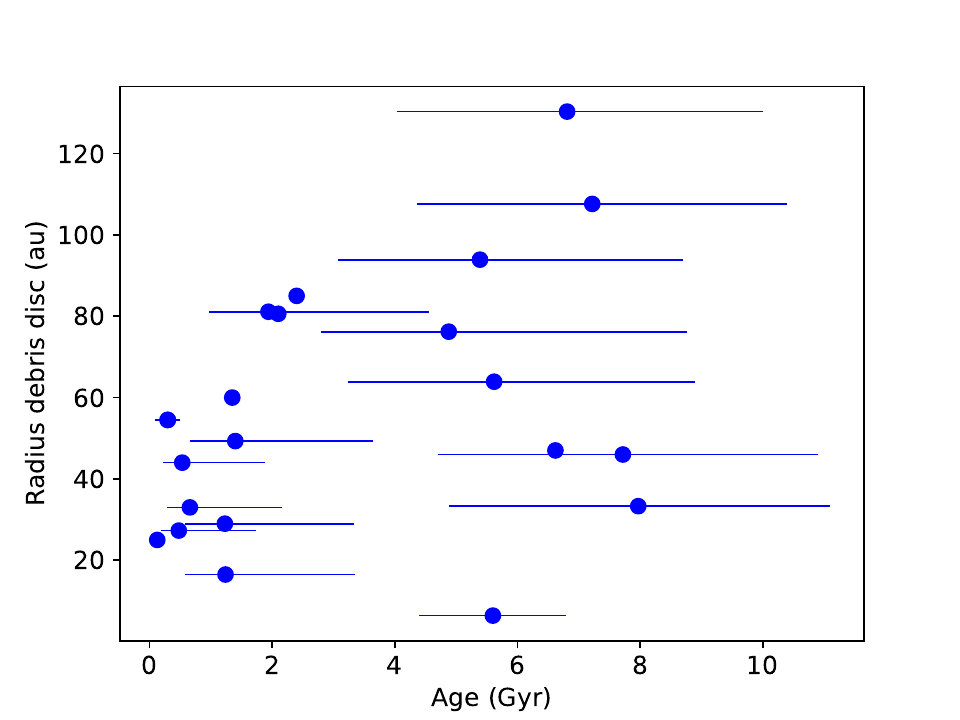}}    
     \caption{Extended debris disc radii vs. space velocity and age. Left panel: Extended debris disc radii as a function of $U_{\mathrm{rel}}$. The blue circles represent debris discs with ages less than 2\,Gyr and the yellow triangles ages greater than 2\,Gyr. The symbol sizes illustrate the relative heliocentric distances.  Right panel: Extended debris disc radii and as a function of stellar age.}
        \label{uvel_R}  
\end{figure*}
Debris discs for which extended emission has been detected can be used as probes of the influence exerted by the ISM. Such an influence would manifest itself in the presence of correlations between parameters that define the debris disc morphology and the magnitude and direction of the host star proper motion with respect to the ISM. To characterise the morphology of the extended discs in our sample (see Table~\ref{table:ext}), we considered their PA, their disc projected ellipticity, $\epsilon$, and their disc radius. Complementing this, we derived the host star velocity vectors in the LIC frame, defined by the angle of $v_{\mathrm{star\_LIC}}$  with the sky perpendicular direction, $\rho$, and by the angle of the sky projection of $v_{\mathrm{star\_LIC}}$ with respect to the east direction of the sky, $\theta$ (see Sect. 2 for further details). 

We investigated whether the radii of the debris discs exhibit a dependence on the space velocity $U_{\mathrm{rel}}$. The left panel of Fig.~\ref{uvel_R} displays the disc radii estimates based on \citet{Marshall2021} as a function of $U_{\mathrm{rel}}$. These authors performed a homogeneous analysis of 95 debris discs observed by \textit{Herschel} that included their SEDs and stellar luminosities. The left panel of Fig.~\ref{uvel_R} shows that the dispersion of debris disc radii increases as $U_{\mathrm{rel}}$ goes from negative to positive (and greater) values. Consequently, the maximum radius of the debris discs for a certain space velocity also increases, an effect that is mostly present for older stars (age \,>\,2\,Gyr). Higher values of $|U|$ could affect the eccentricities of the dust particles, leading to more elongated debris disc shapes and larger apparent radii \citep{Maness2009}. Figure~\ref{uvel_R} displays an additional trend. As $U_{\mathrm{rel}}$ goes from negative to positive and higher values, the minimum radius of the debris discs for a certain space velocity becomes smaller. As mentioned above, higher values of $|U|$ may lead to more elongated and larger debris discs, causing the discs to appear fainter towards their boundaries. For small and thin discs, the eroded edges can be lost (becoming part of the ISM). If confirmed, this behaviour would contradict the assumption that smaller discs are more deeply embedded inside the stellar heliosphere and therefore they should be more resistant against the influence of the LISM. Some caution should be exercised when analysing this negative trend in small debris discs, as the two debris discs with smaller radii are also the nearest, which suggests a potential bias related to the distance to these stars. Additional factors that may affect the disc morphology and the radius derivation include the instrument resolution and the properties of the system, such as the disc optical depth and the inclination. The left panel of Fig.~\ref{uvel_R} also shows that, in younger stars (age \,<\,2\,Gyr), debris discs are smaller and decrease in size with $U_{\mathrm{rel}}$ faster, on average, than in older stars. This observation is consistent with the stronger decline in debris disc occurrence rates with $|U_{\mathrm{rel}}|$ for younger targets reported in the previous subsection.

The right panel of Fig.~\ref{uvel_R} shows the debris disc radii as a function of stellar age. The trend for larger disc radii resembles that in the left panel, though the significant uncertainties hinder a more precise comparison. \citet{Eiroa2013} also found an apparent correlation between debris disc radii and stellar ages in the DUNES sample and suggested that it might be a sign of dynamical inward-out stirring of debris discs. This effect could contribute to the increase in debris disc radii with age and therefore with $|U_{\mathrm{rel}}|$, as older stars are associated with a larger dispersion of $|U|$ values. However, the opposite trend present for smaller discs is stronger with $|U_{\mathrm{rel}}|$ than with age, further supporting that it may be caused by the interaction with the LISM. In summary, Fig.~\ref{uvel_R} probably represents the combined influence of the LISM and the system's age.

The resolved and unresolved debris discs in our sample have the same distribution with respect to the space velocity $U$. Our resolved debris discs have on average larger dust fractional luminosities than our unresolved debris discs, but, except for the two largest discs, their dust fractional luminosities vary in a range between one and two orders of magnitude for debris discs of the same size. We can assume that this variation is also present in unresolved discs, and therefore that their being unresolved is associated with not only their disc size but also their fractional luminosities and distances. This explains why we do not see a link between the fact that a disc is resolved or not and their stellar space velocity. 
\begin{figure}
\resizebox{\hsize}{!}
        {\includegraphics[width=\hsize]{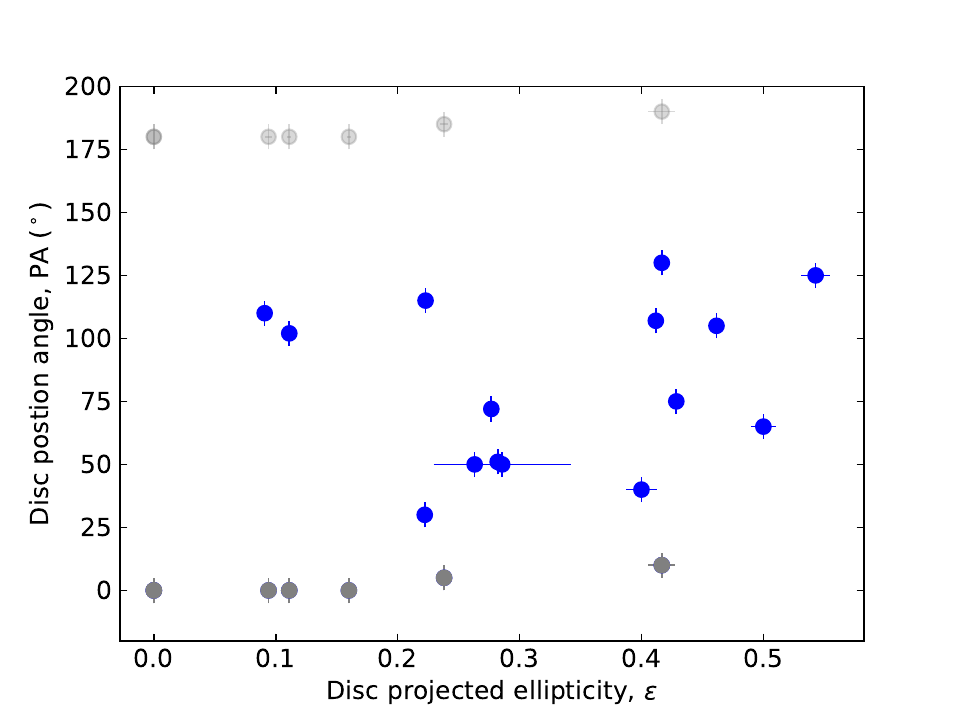}}
        \caption{PAs of the debris discs as a function of the discs projected ellipticities, $\epsilon$. Discs with PA\,${\leq 10\degr}$ are represented with grey dots because, considering their uncertainty ($\pm 5\degr$), their actual value could be about $180\degr$. To illustrate the effect on the trend, the corresponding values around  $180\degr$ are also shown in light grey.}
    \label{ellip_pa}
\end{figure}

Figure~\ref{ellip_pa} shows the relation between the discs projected ellipticities, $\epsilon$ and their PA. We see a weak indication that discs with small PA, that is, closer to the north--south direction, have smaller ellipticities, which would be consistent with a scenario in which a common external agent, possibly the LISM, influences the orientation and inclination of these debris discs. However, discs with $\mathrm{PA} \leq 10$, shown as grey dots, which might have real values around $180\degr$ given their uncertainty of $\pm 5\degr$, could eliminate the observed trend.

\begin{figure}
\resizebox{\hsize}{!}
        {\includegraphics[width=\hsize]{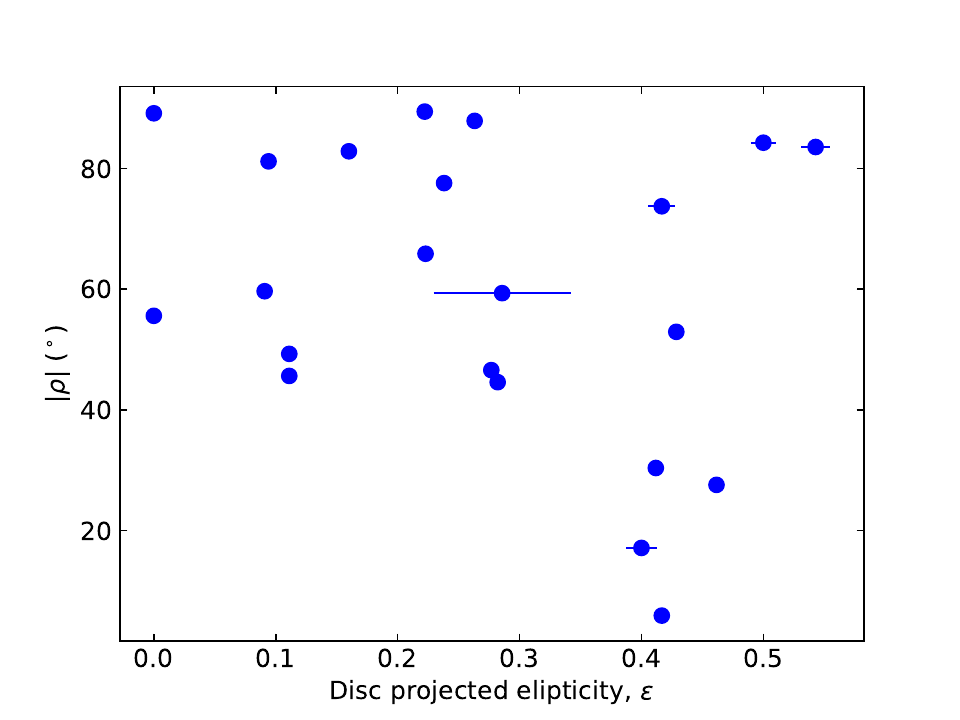}}
        \caption{Angles $|\rho|$ of the relative velocities in the LISM of the host stars as a function of the discs' projected ellipticities, $\epsilon$.}\label{ellip_ro}
\end{figure}
Figure~\ref{ellip_ro} represents the disc projected ellipticities, $\epsilon$, as a function of $|\rho|$. Host targets with smaller values of $|\rho|$, that is, with radial velocity components greater than the respective on-sky components, have debris discs with greater projected ellipticity, $\epsilon$. One potential interpretation is that, at greater relative radial velocities, the LISM has an increased ability to tilt debris discs more steeply in relation to the sky plane. Two objects exhibit $\epsilon \geq 0.5$ with $|\rho|$ values near 90$\degr$. Given that their relative velocities within the LISM are on the plane of the sky, their elevated $\epsilon$ values likely result from an initial near edge-on disc orientation as perceived by the observer. The lower left region of Fig.~\ref{ellip_ro}, which would correspond to debris discs with low ellipticity and a radial $v_{\mathrm{star\_LIC}}$, is empty. This is expected if debris discs are tilted to be parallel to their $v_{\mathrm{star\_LIC}}$. 

\begin{figure}
\resizebox{\hsize}{!}
    {\includegraphics{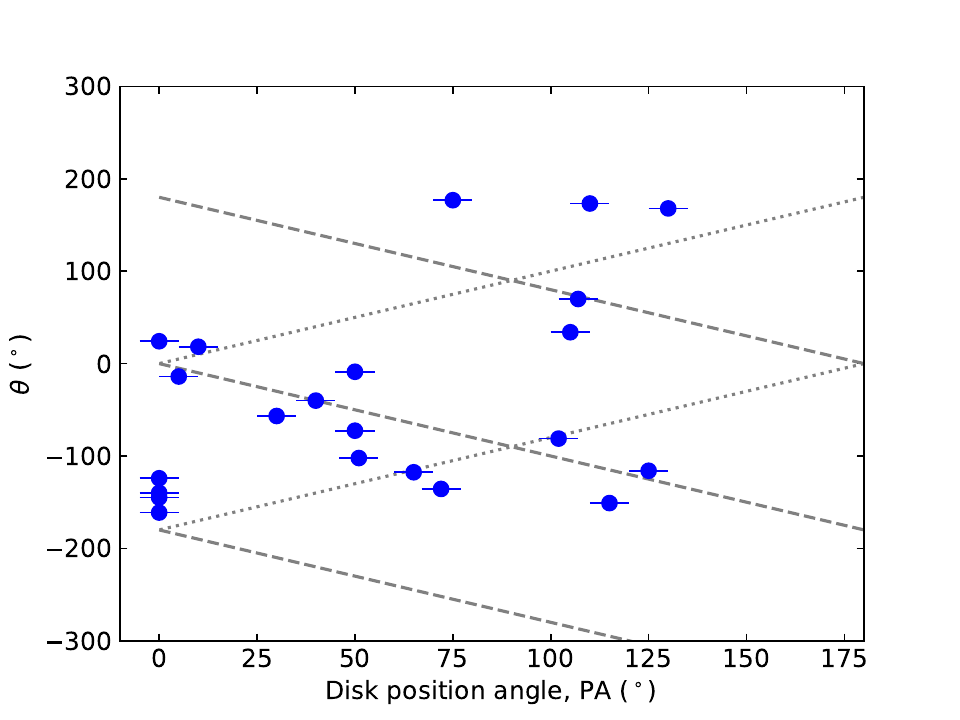}}
        \caption{Angles, $\theta,$ of the host stars' relative velocities in the LISM as a function of the debris discs PAs. The dashed lines show the points where the difference between $\theta$ and PA is $90\degr$ for a disc with positive inclinations, and the dotted lines display the equivalent relation for discs with a negative inclination. This broad correlations have been extracted from the simulations made by \citet{Maness2009} when the ISM flow is on the same plane as the disc.}\label{pa_theta}
\end{figure}

The possible influence of the LISM on the debris disc morphology is further explored in Fig.~\ref{pa_theta}, which displays $\theta$, the angle on the plane of the sky of the star relative velocity, measured with respect to the east direction, as a function of the debris disc PA. The dashed lines show the points in which the difference between $\theta$ and PA is $90\degr$ for discs with positive inclinations, and the dotted lines display the equivalent relation for discs with negative inclinations, as derived from the simulations in \citet{Maness2009} when the ISM flow is in the same plane as the disc. These authors and \citet{Marzari2011} predict that perturbations induced by the ISM generate asymmetries: debris discs tend to be elongated in the direction perpendicular to the ISM flow when it is on the same plane as the disc. \citet{Marzari2011} showed that these effects are apparent in thin discs with low optical depth, $\tau \approx 10^{-6}$, which is also the case for the debris discs in our sample. In Fig.~\ref{pa_theta}, we can see that although there is some dispersion, the symbols tend to group around these lines. 

\begin{figure*}
\resizebox{\hsize}{!}
    {\includegraphics{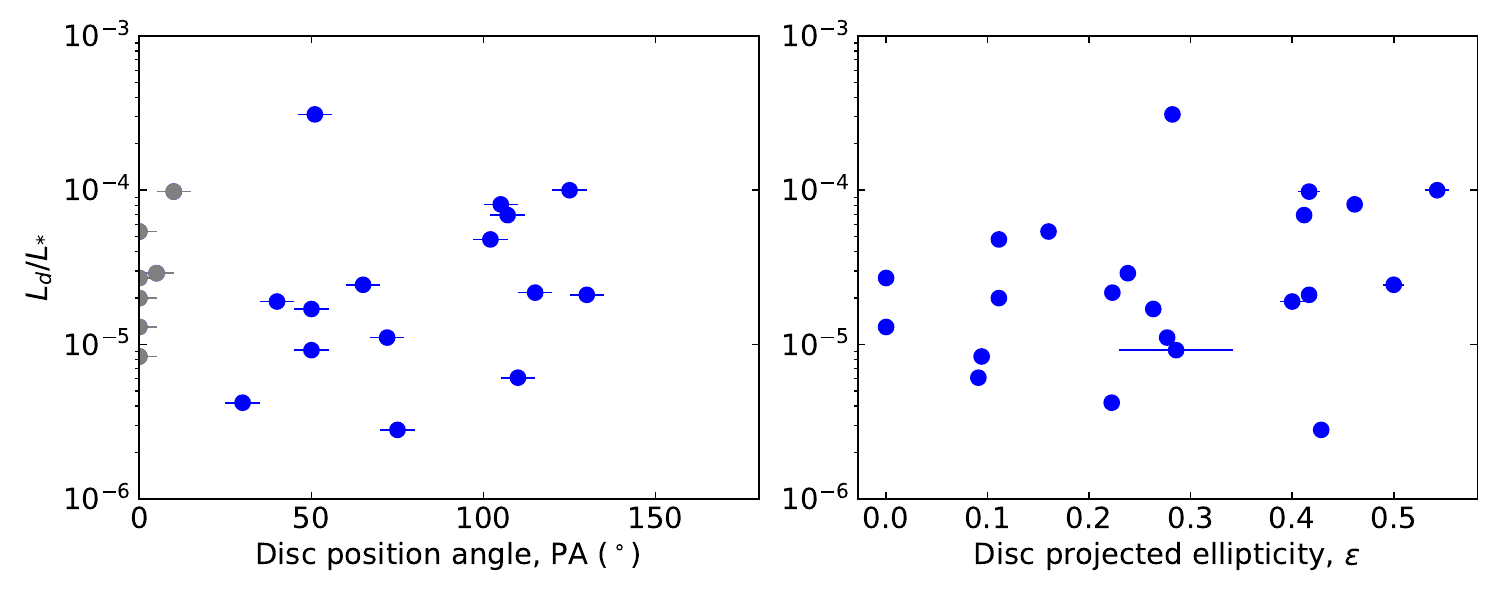}}
        \caption{Debris disc fractional luminosities as a function of the debris disc PAs (left panel) and $\epsilon$ (right panel). Discs with PA\,${\leq 10}$ are represented with grey dots because, considering their uncertainty ($\pm 5\degr$), their actual value could be approximately $180\degr$. }
        \label{pa_dust}
\end{figure*}

If the debris discs in our neighbourhood are influenced by the LISM, not only should their disc morphologies be affected, but also their dust contents, dust distribution, and overall mass and brightness. Indeed, as shown in the left panel of Fig.~\ref{pa_dust}, there is a tentative positive trend in the fractional luminosities of the debris discs versus their PA. The fractional luminosities increase with PA, from debris discs orientated close to the north--south direction to a maximum for debris discs orientated in the sky with a PA of $\approx 150\degr$. Discs with PA\,${\leq 10\degr}$ have been excluded from this analysis because their actual value could also be about $180\degr$, given their uncertainty ($\pm 5\degr$).  The object exhibiting the highest disc fractional luminosity deviates from the expected trend. This observation aligns with the prediction made by \citet{Marzari2011}, suggesting that the optical depth of the disc plays a crucial role in making debris discs vulnerable to the influence of the ISM. In summary, Fig.~\ref{pa_dust} tentatively infers that the LISM affects the dust content depending on the disc orientation, although a larger number of targets would be required to confirm the apparent trend. In the right panel, we do not see any correlation of the dust fractional luminosity with $\epsilon$.
\section{Discussion}
Debris disc occurrence, geometry, and morphology have been found to depend on the conditions during stellar formation, the presence of planets and other bodies, and the stellar spectral type and age \citep[e.g.][]{Nordstrom2004,Wyatt2012,Sierchio2014,Marshall2014b,Eiroa2013,Montesinos2016,
Moor2016,Meshkat2017,Goldman2018}. The contribution of all these effects may blur the influence of the LISM, and so disentangling its contribution becomes a challenging task. In this context, statistical analyses of defined samples of stars that are embedded in similar LISM conditions can provide useful indications that may not be apparent when considering objects in isolation. That is why our study was based on the DUNES and DEBRIS debris discs, a unique set of nearby objects located in the well-studied LIC region. An additional advantage of our sample is that it consists of debris discs with low optical depths, which are much more susceptible to being influenced by the LISM than optically thick discs. \cite{Marzari2011} modelled the ISM influence on debris discs and concluded that it was negligible when the disc optical depths were greater than 10$^{-3}$. On the other hand, for discs with optical depths smaller than 10$^{-3}$, the models showed that peculiar asymmetric patterns appeared in the density profile of the discs. 

We propose that the observed decrease in the debris disc occurrence rate with $|U|$ can be attributed to the removal of dust grains caused by the interaction with the LISM. As found by \citet{Nordstrom2004}, the dispersion of space velocity values increases slowly with stellar age, and therefore the decrease in the occurrence rates may partially be due to the fact that stars with larger $|U|$ tend to be older. However, in Sect.~\ref{section:occurrence} we provide evidence that the anti-correlation of the occurrence rate with increasing $|U|$ is still present even after considering the stellar age. This conclusion still depends on the stellar age determinations, which for many stars may be rather uncertain. Therefore, the connection between $|U|$, stellar age, and debris disc occurrence rates should be revisited as more accurate stellar age determinations are available. Another possibility is to further analyse this correlation for stars in stellar clusters, young moving groups, and stellar associations, which are better laboratories for a more accurate determination of stellar ages and provide a variety of environments. 

Debris discs are created by the replenishment of dust through collisions of bodies in the stellar system. Their collisional dust depletion and replenishment times are both relatively short, in the range 0.01-1 Myr \citep[e.g.][]{Arty1997, Ahmic2009}. The evolution of a debris disc is determined by the timescales of their collisional cascades and the forces that are exerted on the dust particles \citep[e.g.][]{Wyatt2008, Hughes2018}. As the population of destructible planetesimals is eroded, dust production from collisional cascades decreases over time, causing the expected depletion of the debris disc dust with stellar age. However, the timescales for the effect of forces directly exerted on the dust grains, such as those by the LISM, can have much shorter timescales. Therefore, in today’s observations we should see the superposition of the long-term effect of the changing population of parent bodies and large grains that generate the collisional cascades, for which the age is relevant, and the short-term effect of the LISM. 

The influence of the ISM varies as a function of time as stars travel in the galaxy and cross different ISM density regions. All stars in our sample are currently in the Local Bubble, but have experienced different conditions since their birth. The Sun, in particular, has orbited the Galactic centre slightly more than 20 times over the course of its 4.6 billion years of existence. Nevertheless, because of the short dust replenishment characteristic times, we are observing relatively new dust, so the debris disc occurrence rate and morphologies are affected by the LISM conditions encountered by the stars in the last few million years. In particular, the Sun entered the Local Bubble around 5\,Myr ago. Likewise, most of the stars in our sample have spent at least 2\,Myr in the Local Bubble; we can thus assume that the observed debris discs have developed in this environment. On the other hand, the small group of stars with $|U| \gtrapprox 50$\,km s$^{-1}$ could have entered the Local Bubble about 1\,Myr ago, having previously moved through a much denser ISM. Outside the Local Bubble, their debris discs could have been more strongly affected by the ISM, further reducing the likelihood of the presence of debris discs, as we have observed. In the mechanism proposed by \citet{Scherer2000} and considered in \citet{Maness2009}, the force exerted by the neutral ISM gas on the Solar System dust is such that 10\,$\mu$m particles at 40\,au become unbound after 1\,Myr. The timescale for this mechanism to operate is therefore compatible with the time that most stars of our sample have spent in the Local Bubble. In this regard, it is important to emphasise that \textit{Herschel} data are essential because, although millimetre-resolved imaging of discs (e.g. from  ALMA) provide higher resolution for the debris disc characterisation, they trace larger dust grains that are longer lived and consequently carry information of effects that have longer timescales than the duration our sample stars spend in the Local Bubble.  

Although the theoretical models for debris disc evolution establish a clear functional dependence with stellar age, the comparison with the observations still leaves unanswered questions. Based on the fractional disc luminosity of the debris discs observed by \citet{Matra2018b}, \textit{Spitzer} and \textit{Herschel}, \citet{Najita2022} stated that debris disc detection fell along a wide swath in fractional luminosity and declined with time, and that current data suggested that the frequency of cold debris discs was roughly constant at approximately 25\% for stellar ages of 50\,Myr to 10\,Gyr. These authors proposed that rings of solids at a distance of 30--40\,au offered a way to resolve a long-standing disconnect between detailed evolutionary models of debris discs and their observed properties. Based on our results, we suggest that, to reconcile the observed and theoretical evolution of the debris discs, the environment should also be taken in consideration.

In our analysis, we used the disc projections in the sky to study the effect of the ISM. In contrast with previous work, we assumed that the debris disc ellipticities seen in our \textit{Herschel}/PACS observations result from a combination of the disc inclination and the sculpting by the LISM. Typically, the ellipticity observed in debris discs has been attributed solely to their inclination. This assumption has been supported by the lack of asymmetries in many debris discs, as observed by ALMA. However, most debris discs are marginally resolved at millimetre wavelengths, and therefore the significance of these observations is resolution- and sensitivity-limited. Moreover, greater asymmetries are observed when probing smaller dust grains. For example, \citet{MacGregor2019} noted that the debris disc around HD 15115 was fairly symmetric in ALMA observations, contrasting with the significant east-west asymmetry seen in scattered light with the Hubble Space Telescope \citep{Kalas2007}. Consequently, \citet{MacGregor2019} concluded that this asymmetric feature most likely resulted from a mechanism that only affects small grains.

The ISM could also affect the thickness of the discs. One may assume that, if a disc is being ablated by the ISM and by planets, those belts with high $|U|$ would be narrower than others, all else being equal. However, as shown in \citet{Marzari2011}, under certain circumstances defined by the value and direction of $U$, the optical depth, and the dust grain size, particles can build large eccentricities and inclinations, increasing the debris disc thickness when seen edge-on. Unfortunately, most of our extended discs are unresolved in their vertical direction \citep{Marshall2021}, preventing us from studying this question further. Recent advances have enabled the exploration of vertical dust distribution in debris discs. Specifically, \cite{Terrill2023} introduced a novel approach to determining the radial and vertical structures of 16 highly inclined debris discs as observed by ALMA (one of them, HIP 7978, is also in our sample). 

New influencing factors have recently been reported. For example, \citet{Bertini2023} found that in the period between the last 5 Myrs and the next 2 Myrs, 90\% of the analysed systems would experience at least one close flyby and that these events may fundamentally impact the evolution of debris discs. Impacts are more frequent in the early stellar formation phase and when the star belongs to a cluster, a young moving group, or a stellar association. The higher the number density, the more likely a close encounter will take place. However, field stars (such as the Sun) are much less likely to be affected by tidal disruption, and the impact of the LISM is the main external `force' to consider. 

As our debris disc sample is based on \textit{Herschel} observations in the far-infrared, we are mainly probing the influence of the environment at distances from the stars that range from the Kuiper-Belt distance to the Sun up to a few hundred au. These far regions may be more likely affected by the environment than warmer regions closer to the star. Extending our analysis to near- and mid-infrared observations of debris discs will be a next step in studying the influence of the ISM on dust grains under regimes different from those of the external parts of the disc. 

\section{Conclusions}
We have studied the influence of the ISM on debris disc occurrence rates and debris disc morphologies. To perform this analysis, we used the results of the \textit{Herschel} Space Observatory DUNES and DEBRIS surveys, in particular the images of 295 nearby FGK dwarf stars at 100\,$\mu$m and 160\,$\mu$m \citep{Eiroa2013, Matthews2010, Montesinos2016, Sibthorpe2018}. Most of the debris discs observed in the sample have low optical depths, which makes them more likely to be affected by the LISM compared to optically thick discs, as predicted by \citet{Marzari2011}. Given that these stars are located within the LIC, an area with well-studied properties, we can infer that their debris discs encounter similar environmental conditions, thereby confirming our statistical approach. This allows us to use the stellar space velocity, represented by its components ($U$, $V$, and $W$), as an indicator of the forces that can act on the debris disc dust grains when they interact with the ISM, all other environmental factors being equal. These forces introduce perturbations in the orbits of the dust particles, affecting the disc shapes and causing some grains to migrate towards the star or be ejected out of the system. 
\par
Our analysis of the debris disc occurrence rates shows that:
   \begin{enumerate}
      \item The percentage of sources with debris discs reaches a maximum of about 25\% for stars with low values of $|U_{\mathrm{rel}}|$ and decreases monotonically for larger $|U_{\mathrm{rel}}|$ values down to the 10\% level. A similar behaviour is seen for the disc fractional luminosities averaged per bin. This supports the notion that stars with greater space velocities have a higher probability of losing their circumstellar dust.  
      \item The decrease in the occurrence rates with $|U_{\mathrm{rel}}|$ is more significant in younger stars (age < 3 Gyr). 
      \item The dependence of the disc fractional luminosity on $|U_{\mathrm{rel}}|$ does not disappear, even after accounting for the reported higher dispersion of $U$ values with age. 
      \item The dependence of debris disc occurrence rates on the $V_{\mathrm{rel}}$ velocity component is less apparent than with $U_{\mathrm{rel}}$, and is not present in association with the $W_{\mathrm{rel}}$ velocity component when $W_{\mathrm{rel}}$ < 30 km\,s$^{-1}$. For a small number of stars with $W_{\mathrm{rel}}$ greater than this value, the occurrence rate is significantly lower. For both $V_{\mathrm{rel}}$ and $W_{\mathrm{rel}}$, the fractional luminosities of the discs averaged per bin follow a pattern similar to that of the occurrence rates.
        \item The occurrence rate of debris discs is marginally lower for stars at distances from 5 to 10 pc than for stars at larger distances of up to 18 pc. This could be explained by assuming that the LISM is denser within $\approx$10 pc, a scenario that is consistent with the model proposed by \citet{Gry2014}. 
        \item  The probability density distribution as a function of $|U_{\mathrm{rel}}|$ of sources with planets and debris discs appears to be narrower than the distribution of sources with planets without debris discs. A possible explanation is that stars with planets and high space velocities might have lost part or most of their circumstellar dust as a consequence of the interaction with the LISM. This possible dust depletion could have implications when looking for correlations between the properties of debris discs and the presence of planets.
   \end{enumerate}

The influence of the LISM on debris discs is further substantiated by our findings on the morphologies of extended discs:
    \begin{enumerate}
        \item As $U_{\mathrm{rel}}$ goes from negative to positive (and greater) values, a positive trend is observed in the size of the largest debris discs, notably in older stars (age \,>\,2\,Gyr), whereas a negative trend is apparent in the size of the smaller debris discs, predominantly in younger stars (age \,< \,2\,Gyr).
        \item The disc projected ellipticities, $\epsilon$, and the disc PAs show a tentative correlation, suggesting that a common agent influences the orientation and inclination of the debris discs. 
        \item In objects with a small angle, $\rho,$ between the stellar relative space velocity and the radial direction, the debris discs show larger projected ellipticities. 
        \item The debris disc PAs tend to be perpendicular to the on-sky component of their relative space velocities, which is consistent with the theoretical predictions by \citet{Scherer2000}, \citet{Maness2009}, and \citet{Marzari2011}. 
        \item The fractional luminosities of the debris discs appear to be correlated with their PAs, suggesting that the effect of the ISM on the dust content depends on the disc orientation. 
    \end {enumerate}

The statistical significance of our results is limited because the number of debris discs in our sample is relatively small and debris discs are influenced by a large number of factors associated with their formation and evolution. However, we have found various indications that, taken together, reinforce the hypothesis that the ISM impacts the occurrence rates, dust content, and morphologies of debris discs. Theoretical models and the study of individual debris discs have already highlighted the importance of the environment. This first systematic observational study aligns with those results and indicates that the influence of the ISM requires further investigation and consideration in future debris disc research, particularly when the discs are optically thin.
\section*{Data availability}
Table A.1 is only available in electronic form at the CDS via anonymous ftp to cdsarc.u-strasbg.fr (130.79.128.5) or via \href{http://cdsweb.u-strasbg.fr/cgi-bin/qcat?J/A+A/}{http://cdsweb.u-strasbg.fr/cgi-bin/qcat?J/A+A/}. 

\begin{acknowledgements}
We thank an anonymous referee for their insightful comments and suggestions, which have improved the quality of this article. This work has made use of data from \textit{Herschel}, an ESA space observatory with science instruments provided by European-led Principal Investigator consortia and with important participation from NASA.
This work has made use of data from the European Space Agency (ESA) mission
{\it Gaia} (\url{https://www.cosmos.esa.int/gaia}), processed by the {\it Gaia}
Data Processing and Analysis Consortium (DPAC,
\url{https://www.cosmos.esa.int/web/gaia/dpac/consortium}). Funding for the DPAC
has been provided by national institutions, in particular the institutions
participating in the {\it Gaia} Multilateral Agreement.
      This research has made use of the VizieR catalogue access tool, CDS, Strasbourg, France. 
      The original description of the VizieR service was published in A\&AS 143, 23
      This research has made use of NASA’s Astrophysics Data System Bibliographic Services. The figures in this article have been produced using Matplotlib \citep{Hunter:2007}. This research has made use of data obtained from or tools provided by the portal exoplanet.eu of The Extrasolar Planets Encyclopaedia. C.dB. acknowledges support from a Beatriz Galindo senior fellowship (BG22/00166) from the Spanish Ministry of Universities. J.P.M. acknowledges support by the National Science and Technology Council of Taiwan under grant NSTC 112-2112-M-001-032-MY3. B.M. is partially funded by grant PID2021-127289-NB-I00 by the Spanish Ministry of Science and Innovation/State Agency of Research (MCIN/AEI).
\end{acknowledgements}
%
%
%
\bibliographystyle{aa} 
\bibliography{References_disk_bib} %
\end{document}